\documentclass[a4paper,11pt]{article}
\usepackage{pos}

\title{
    From Theory to Practice:
    Applying Neural Networks to Simulate Real Systems with Sign Problems
}
\ShortTitle{Neural Networks vs Sign Problem}

\usepackage{physics}
\usepackage{siunitx}
\usepackage[super]{nth}
\usepackage{nicefrac}

\usepackage{caption}
\usepackage{subcaption}

\usepackage{tikz}
    \usetikzlibrary{calc,arrows,matrix,arrows.meta,shapes,decorations,decorations.shapes}
    \tikzstyle{arrow} = [very thick,->,>=stealth]

\definecolor{fzjblue}{RGB}{2,61,107}
\definecolor{fzjlightblue}{RGB}{173,189,227}
\definecolor{fzjgray}{RGB}{235,235,235}
\definecolor{fzjred}{RGB}{235, 95, 115}
\definecolor{fzjgreen}{RGB}{185, 210, 95} 
\definecolor{fzjyellow}{RGB}{250, 235, 90}
\definecolor{fzjviolet}{RGB}{175, 130, 185}
\definecolor{fzjorange}{RGB}{250, 180, 90}
\definecolor{fzjblack}{RGB}{0,0,0}
\definecolor{fzjwhite}{RGB}{255,255,255}

\newcommand{\Obs}{\ensuremath{\mathcal{O}}}
\newcommand{\Z}{\ensuremath{\mathrm{Z}}}

\DeclareMathOperator{\im}{i}
\newcommand{\Reals}{\mathbb{R}}
\newcommand{\Complexes}{\mathbb{C}}
\newcommand{\DD}[1]{\mathcal{D}\left[#1\right]}
\newcommand{\Ncfg}{\ensuremath{\mathrm{N}_{\mathrm{cfg}}}}

\newcommand{\NN}[1]{\mathrm{NN}\left[#1\right]}

\tikzstyle{startstop} = [rectangle, rounded corners, minimum width=3cm, minimum height=1cm,text centered, draw=black, fill=fzjred]
\tikzstyle{process} = [rectangle, minimum width=3cm, minimum height=1cm, text centered, draw=black, fill=fzjblue!30]
\tikzstyle{decision} = [rectangle, rounded corners, minimum width=3cm, minimum height=1cm, text centered, draw=black, fill=fzjgreen]

\newcommand{\jsc}{
    JARA \& J\"{u}lich Supercomputing Center,
    Forschungszentrum J\"{u}lich, 54245 J\"{u}lich, Germany
}
\newcommand{\ias}{
    Institute for Advanced Simulation,
    Forschungszentrum J\"{u}lich, 54245 J\"{u}lich, Germany
}
\newcommand{\casa}{
    Center for Advanced Simulation and Analytics (CASA),
    Forschungszentrum Jülich, 52425 J\"{u}lich, Germany
}
\newcommand{\bonn}{
    Helmholtz-Institut f\"{u}r Strahlen- und Kernphysik,
    Rheinische Friedrich-Wilhelms-Universit\"{a}t Bonn, 53115 Bonn, Germany
}
\newcommand{\oxford}{
    Department of Physics, Condensed Matter Physics sub-department, Clarendon Laboratory, 
    University of Oxford, Parks Road, Oxford OX1 3PU, United Kingdom
}

\affiliation[a]{\jsc}
\affiliation[b]{\ias}
\affiliation[c]{\bonn}
\affiliation[d]{\casa}
\affiliation[e]{\oxford}

\author*[a,b,c,d]{Marcel Rodekamp}
\emailAdd{m.rodekamp@fz-juelich.de}

\author[a,b,d]{Evan Berkowitz}
\emailAdd{e.berkowitz@fz-juelich.de}

\author[e]{Maria Dincă}
\emailAdd{maria.dinca0501@gmail.com}

\author[b,c,d]{Christoph Gäntgen}
\emailAdd{c.gaentgen@fz-juelich.de}

\author[a,b,c,d]{Stefan Krieg}
\emailAdd{s.krieg@fz-juelich.de}

\author[b,c,d]{Thomas Luu}
\emailAdd{t.luu@fz-juelich.de}

\abstract{
    The numerical sign problem poses a seemingly insurmountable barrier to the simulation of many fascinating systems.
    We apply neural networks to deform the region of integration, mitigating the sign problem of systems with strongly correlated electrons.
    In this talk we present our latest architectural developments as applied to contour deformation.
    We also demonstrate its applicability to real systems, namely perylene.
}

\date{\today}

\begin{document}
\maketitle

\section{Introduction}
The challenge of the computational sign problem poses obstacles for effective importance sampling when dealing with complex-valued distributions in Markov-Chain Monte-Carlo algorithms,
like Hybrid Monte Carlo (HMC). 
This challenge is pervasive when sampling from the configuration space of many physical systems, 
including but not limited to lattice QCD at finite baryon chemical potential, 
doped condensed matter systems in equilibrium, 
and the real-time evolution of quantum systems.
A method to substantially alleviate the sign problem involves deforming the original (real) manifold of integration into a sign improved (complex) one~\cite{Kashiwa:2018vxr,alexandru2020complex,Detmold:2020ncp,Detmold:2021ulb}. 
Formal developments have spurred exploration into leveraging Lefschetz thimbles~\cite{Lefschetz1921,PhysRevD.93.014504,Cristoforetti:2014gsa,Cristoforetti:2013wha,Mukherjee:2013aga,Kanazawa:2014qma,Tanizaki:2016lta}
—these are high-dimensional counterparts to contours of steepest descent and can be located through holomorphic flow. 
However, pinpointing the location of each relevant thimble's saddle point, or critical point, 
and determining the relevant sampling `direction' around these points is prohibitive. 
An alternative approach is to employ neural networks to learn the mapping from an initial manifold to a beneficial one, 
including one that approximates the contributing thimbles to the integral~\cite{PhysRevD.96.094505,Mori:2017nwj,leveragingML,PhysRevB.106.125139}.

In this work, we explore the use of this method for perylene, $C_{20}H_{12}$~\cite{botoshanskyCompleteDescriptionPolymorphic2003}.
The molecule is (almost) planar and $\mathrm{sp}^2$-hybridized making it a perfect target for the Hubbard model~\cite{camermanCrystalMolecularStructure1953}.
We show some preleminary results for a chemical potential (doping) scan.
Practical applications range from organic semiconductors~\cite{dodabalapur1996molecular},
over organic light emitting diodes (OLEDs)~\cite{sato1998operation},
to organic solar cells~\cite{tang1986two}.
Perylene also falls under the class of so-called polycyclic aromatic hydrocarbon (PAH) molecules thought to be ubiquitous in interstellar gases and nebulae~\cite{salama_2008}.  
A detailed accounting of electronic properties of PAHs, like perylene, could help constrain interstellar gas models~\cite{Aigen2023}.
 
\section{Formalism}
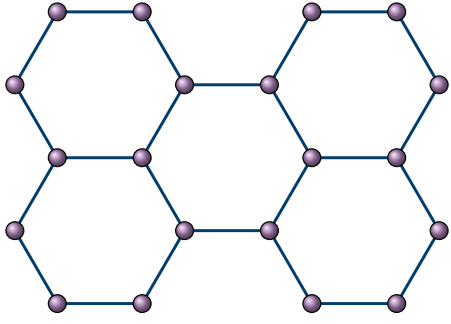
\begin{figure}
    \centering
    \resizebox{0.4\textwidth}{!}{
        \begin{tikzpicture}
            \coordinate (N01) at (0,0);
            \coordinate (N02) at (1,0);
            \coordinate (N03) at (3,0);
            \coordinate (N04) at (4,0);
            \coordinate (N05) at (4.5,-0.8660254037844386);
            \coordinate (N06) at (2.5,-0.8660254037844386);
            \coordinate (N07) at (1.5,-0.8660254037844386);
            \coordinate (N08) at (-0.5,-0.8660254037844386);
            \coordinate (N09) at (0,-1.73205080756888);
            \coordinate (N10) at (1,-1.73205080756888);
            \coordinate (N11) at (3,-1.73205080756888);
            \coordinate (N12) at (4,-1.73205080756888);
            \coordinate (N13) at (4.5,-2.59807621135332);
            \coordinate (N14) at (2.5,-2.59807621135332);
            \coordinate (N15) at (1.5,-2.59807621135332);
            \coordinate (N16) at (-0.5,-2.59807621135332);
            \coordinate (N17) at (0,-3.46410161513775);
            \coordinate (N18) at (1,-3.46410161513775);
            \coordinate (N19) at (3,-3.46410161513775);
            \coordinate (N20) at (4,-3.46410161513775);
        
            \draw[fzjblue,line width = 1px] (N01) -- (N02);
            \draw[fzjblue,line width = 1px] (N02) -- (N07);
            \draw[fzjblue,line width = 1px] (N07) -- (N06);
            \draw[fzjblue,line width = 1px] (N06) -- (N03);
            \draw[fzjblue,line width = 1px] (N03) -- (N04);
            \draw[fzjblue,line width = 1px] (N04) -- (N05);
            \draw[fzjblue,line width = 1px] (N05) -- (N12);
            \draw[fzjblue,line width = 1px] (N12) -- (N13);
            \draw[fzjblue,line width = 1px] (N13) -- (N20);
            \draw[fzjblue,line width = 1px] (N20) -- (N19);
            \draw[fzjblue,line width = 1px] (N19) -- (N14);
            \draw[fzjblue,line width = 1px] (N14) -- (N15);
            \draw[fzjblue,line width = 1px] (N15) -- (N18);
            \draw[fzjblue,line width = 1px] (N18) -- (N17);
            \draw[fzjblue,line width = 1px] (N17) -- (N16);
            \draw[fzjblue,line width = 1px] (N16) -- (N09);
            \draw[fzjblue,line width = 1px] (N09) -- (N08);
            \draw[fzjblue,line width = 1px] (N08) -- (N01);
            \draw[fzjblue,line width = 1px] (N09) -- (N10);
            \draw[fzjblue,line width = 1px] (N10) -- (N07);
            \draw[fzjblue,line width = 1px] (N10) -- (N15);
            \draw[fzjblue,line width = 1px] (N12) -- (N11);
            \draw[fzjblue,line width = 1px] (N11) -- (N06);
            \draw[fzjblue,line width = 1px] (N11) -- (N14);
        
            \draw[ball color=fzjviolet] (N01) circle (3pt);
            \draw[ball color=fzjviolet] (N02) circle (3pt);
            \draw[ball color=fzjviolet] (N03) circle (3pt);
            \draw[ball color=fzjviolet] (N04) circle (3pt);
            \draw[ball color=fzjviolet] (N05) circle (3pt);
            \draw[ball color=fzjviolet] (N06) circle (3pt);
            \draw[ball color=fzjviolet] (N07) circle (3pt);
            \draw[ball color=fzjviolet] (N08) circle (3pt);
            \draw[ball color=fzjviolet] (N09) circle (3pt);
            \draw[ball color=fzjviolet] (N10) circle (3pt);
            \draw[ball color=fzjviolet] (N11) circle (3pt);
            \draw[ball color=fzjviolet] (N12) circle (3pt);
            \draw[ball color=fzjviolet] (N13) circle (3pt);
            \draw[ball color=fzjviolet] (N14) circle (3pt);
            \draw[ball color=fzjviolet] (N15) circle (3pt);
            \draw[ball color=fzjviolet] (N16) circle (3pt);
            \draw[ball color=fzjviolet] (N17) circle (3pt);
            \draw[ball color=fzjviolet] (N18) circle (3pt);
            \draw[ball color=fzjviolet] (N19) circle (3pt);
            \draw[ball color=fzjviolet] (N20) circle (3pt);
        \end{tikzpicture}
    }
    \caption{
    Graphical representation of the perylene molecule. 
    The sites represent carbon-ions, while links indicate allowed hopping.
    External hydrogen atoms are not drawn.
    }\label{fig-perylene}
\end{figure}

Given a fixed spatial arrangement of ions $X$, the Hubbard model in particle-hole basis reads~\cite{Hubbard1963}
\begin{align}
    \mathcal{H}\left[K, V, \mu\right]
    =
           - \sum_{x,y\in X} \left( p_x^\dagger K^{xy} p_y - h_x^\dagger K^{xy} h_y \right)
         + \frac{1}{2} \sum_{x,y\in X} \rho_x V^{xy} \rho_y
            + \mu \sum_{x\in X} \rho_x,
    \label{eq:hubbard-hamiltonian}
\end{align}
where the amplitudes in $K$ encode the hopping of fermionic particles $p$ and holes $h$, interactions between these are modelled by the potential $V$,
$
\rho_x = p^{\dagger}_x p_x - h^{\dagger}_x h_x
$ is the net charge per site,
and the chemical potential $\mu$ incentivizes charge.
We restrict our attention to the case where $ K = \kappa \delta_{\langle xy \rangle}$ encodes the structure of perylene, compare figure~\ref{fig-perylene}, with nearest-neighbor hopping, and an on-site interaction $V = U \delta_{xy}$.
In this case, when $\mu=0$, the Hamiltonian in~\eqref{eq:hubbard-hamiltonian} corresponds to the neutral, `half-filled' system.

Trotterizing the thermal trace into $N_t$ timeslices, inserting Grassmannian resolutions of the identity, 
and linearizing the interaction via the Hubbard-Stratonovich transformation~\cite{Hubbard1959} exposes the Hubbard action
\begin{equation}
\begin{aligned}
    S\left[\Phi \,\vert \, K, V, \mu \right]
       =
         \frac{1}{2} \sum_{\substack{x,y\in X\\t \in [0,N_t-1]}} \Phi_{tx} (\delta V^{-1})^{xy} \Phi_{ty}
        -   \log\det{ M\left[\Phi\,\vert\, K,\mu\right] M\left[-\Phi\,\vert\, -K,-\mu\right] },
    \label{eq:hubbard-action}
\end{aligned}
\end{equation}
where $\Phi \in \Reals^{\abs{\Lambda}}$ is the (auxiliary) hubbard field on the spacetime lattice $\Lambda = [0, N_t-1]\otimes X$ and $\delta=\beta/N_t$.
We use the exponential discretization~\cite{Wynen:2018ryx} for the fermion matrices
\begin{equation}
    M\left[\Phi\,\vert\, K,\mu\right]_{x't';xt}
    =
        \delta_{x'x}\delta_{t't}
    -  \left( e^{\delta(K + \mu)} \right)_{x'x} e^{+ i \Phi_{xt}}\delta_{t'(t+1)}
\end{equation}
with antiperiodic boundary conditions in time.
With this we can express the thermal trace as path integral 
\begin{equation}
    \expval{\Obs} = \frac{1}{\Z} \Tr{\Obs e^{-\beta \mathcal{H}}} =  \frac{1}{\Z} \int \DD{\Phi} e^{-S\left[\Phi\right]} \Obs\left[\Phi\right]\label{eq-path-integral}.
\end{equation}
The partition function $\Z$ is the trace/integral without the observable.
On a bipartite lattice we may replace the $-K$ in the holes' fermion matrix with $+K$ leading to a real and positive determinant at vanishing chemical potential. 
When $\mu$ is finite the determinant is complex valued and results in a sign problem.

\section{Method}
As in our previous work~\cite{leveragingML,PhysRevB.106.125139,gantgen2023fermionic}, for a Monte-Carlo algorithm with complex action we can separate the real and imaginary parts, $S=\Re{S}+\im\Im{S}$,
and rewrite the path integral as
\begin{equation}
    \Z = \int \DD{\Phi} e^{-S} = \int \DD{\Phi} e^{-\Re{S}} e^{-\im\Im{S}} \propto \expval{e^{-\im\Im{S}}}_{\Re{S}} \equiv \Sigma
    \label{eq:statistical-power}
\end{equation}
where $\abs{\Sigma}\in[0,1]$ is the statistical power. 
Sampling according to the real part of the action and then applying reweighting,
\begin{align}
        \expval{\Obs}
        &   = \frac{\expval{e^{-\im\Im{S}}\Obs}_{\Re{S}}}{\expval{e^{-\im \Im{S}}}_{\Re{S}}}
        = \frac{1}{\Sigma} \expval{ e^{-\im \Im{S}} \Obs}_{\Re{S}},
        \label{eq:reweighting}
\end{align}
allows the computation of observables.
This procedure fails if the statistical power $\abs{\Sigma}$ cannot be reliably distinguished from zero~\cite{berger2021complex,leveragingML,PhysRevD.93.014504,mori2018lefschetz}.

Combining reweighting with path deformation techniques offers a promising approach to mitigating the sign problem. 
The primary objective is to expand the accessible parameter space for computation, ultimately allowing extrapolations to the continuum limit, $\delta\to0$, and/or the zero-temperature limit, $\beta\to\infty$.
The core idea involves manipulating the integration contour in a way that enhances the statistical power.
One noteworthy method in this context is the use of Lefschetz thimbles, which eliminate fluctuations in the imaginary part of the action.
However, it is essential to acknowledge that these thimbles are notoriously challenging to compute in practical simulations. 
As a result, the focus shifts towards harnessing a more versatile transformation to a sign-optimized manifold, denoted as $\tilde{\mathcal{M}}$. 
Crucially, Cauchy's theorem assures us that expectation values of holomorphic observables remains unchanged under this transformation\footnote{Strictly speaking, this holds when no singularities are crossed when performing the contour deformation, which is the case in our problem.}:
\begin{equation}
\begin{aligned}
    \expval{\Obs} &= \frac{1}{\Z} \int_{\tilde{\mathcal{M}}} \DD{\tilde{\Phi}} e^{-S\left[\tilde{\Phi}\right] } \Obs\left[\tilde{\Phi}\right] \\
                  &= \frac{1}{\Z} \int \DD{\Phi} e^{-S\left[\tilde{\Phi}\left(\Phi\right)\right] + \log\det{J_{\tilde{\Phi}}\left[ \Phi \right] } } \Obs\left[\tilde{\Phi}\left(\Phi\right)\right]
\end{aligned}
\end{equation}

A simple transformation is the mapping to the tangent plane 
$\tilde{\Phi}\left[\Phi\right] = \Phi + \im \phi_0$.
As outlined in~\cite{gantgen2023fermionic} a transcendental equation
\begin{align}
	\label{eq:transcendental}
	\phi_0/\delta = -\frac{U}{N_x}\sum_{k}\tanh\left(\frac{\beta}{2}\left[\epsilon_k+\mu+\phi_0 /\delta\right]\right)
\end{align}
can be used to identify this plane. The sum is over the eigenvalues $\epsilon_k$ of the hopping matrix $K$.
This adds no additional computational cost to the simulation and thus forms our starting point for further transformations.

Based on our earlier machine learning approach~\cite{PhysRevB.106.125139}, we can use a neural network to map to a sign optimized manifold $\tilde{\mathcal{M}}$, i.e. $\tilde{\Phi}\left[\Phi\right] = \NN{\Phi+i\phi_0}$.
We continue to use coupling networks to enable a tractable Jacobian determinant. 
However, we changed the activation function to be holomorphic compared to our earlier approach which simplified the training procedure significantly. 
A single layer thus reads
\begin{equation}
    \mathrm{NN}_l\left[\Psi\right]_{t,x} = 
    \begin{cases}
        e^{s_l\left(\Psi_B\right)} \odot \Psi_A + t_l\left(\Psi_B\right) & (t,x)\in A\\
        \Psi_B & (t,x)\in B
    \end{cases}
\end{equation}
where $s,t: \Complexes^{\nicefrac{\abs{\Lambda}}{2}} \to \Complexes^{\nicefrac{\abs{\Lambda}}{2}},\ \Psi_B \mapsto w' \cdot P\left[ w \cdot \Psi_B +b\right]+b'$ are linear networks; the activation $P\left[\cdot\right]$ is a polynomial of degree 3.
We always pair two of these layers to form a single transformation that allows us to change every value of the input configuration.
Notice, this network is unbounded at $\Psi\to\pm\infty$ and thus escapes Liouville's theorem~\cite{hirose2012complex,lee2022complex}. 
Furthermore, the homology class is completely determined by the asymptotic behavior of the network.
This means that any parametrization of the contour $\tilde{\mathcal{M}}$ preserves the homology class if it approaches a constant asymptotically. 
To achieve this, we implement a projection layer
\begin{equation}
    \mathcal{P}_\sigma\left[\Psi\right]_i = \Re{\Psi}_i + \im \Im{\Psi}_i e^{-\frac{ \abs{\Psi_i}^2 }{\sigma^2} }\label{eq:projection}
\end{equation}
which pushes the imaginary part down towards the real manifold. The parameter $\sigma$ can be chosen arbitrarily. 
Choosing $\sigma\gg 1$ but finite, the projection virtually becomes the identity up to $\order{\sigma^{-2}}$ and the Jacobian remains unchanged
$J_{\tilde{\Phi}}\left[\Phi\right] = \det{\NN{\Phi}} + \order{\sigma^{-2}}$.
Care has to be taken if the configuration value $\NN{\Phi}$ becomes comparable to $\sigma$. 
Naturally, such configuration values have small weights; thus, they practically never appear.

\begin{figure}\centering
    \resizebox{0.6\textwidth}{!}{
\begin{tikzpicture}[node distance = 14em]
    \node[startstop, minimum width = 13em] (start) { $\Phi(\tau), \, \Re{S}_\text{target} \sim e^{-\Re{S}}$};

    \node[process, right of = start, xshift = 1em] (Rkstep) {$\Phi(\tau+\Delta\tau)$};
    \draw[arrow] (start.east) -- node[above,font=\small]{Runge Kutta} (Rkstep.west);

    \node[decision, right of = Rkstep] (Precision) { $\left\vert 1-e^{i\Delta \Im{S} } \right\vert \leq \epsilon$ };
    \draw[arrow] (Rkstep.east) -- (Precision.west);
    \draw[arrow] (Precision.north) -- ($(Precision.north) + (0,1)$) -| node[pos = 0.25,above]{$\Delta\tau \texttt{*=} \delta_\text{attun}$} (start.north);

    \node[decision, below of = Precision, align = center, yshift = 8em] (ReS) { $ \Re{S\left[\Phi(\tau+\Delta\tau)\right]} $\\$<$\\ $ \Re{S}_\text{target} $ };
    \draw[arrow] (Precision.south) -- (ReS.north);
    \draw[arrow] (ReS.west) -| node[above,pos=0.25]{$\tau\texttt{+=}\Delta\tau$} (start.south);  

    \node[decision, below of = Rkstep, yshift = 4em] (store) { $\tau > \tau_\text{min}$ };
    \draw[arrow] (ReS.south) |- (store.east);
    
    \node[process, below of = store, yshift = 10em] (storeExec) {Store Trajectory};
    \draw[arrow] (store.south) -- node[midway,right,font=\small]{True} (storeExec.north);
 
    \node[startstop, left of = store, align = center] (stop) { Next Trajectory };
    \draw[arrow] (store.west) -- (stop.east);

\end{tikzpicture}
}
    \caption{
        Algorithm to generate training data. 
        Each step integrates the flow equation for a time $\Delta\tau$ and checks that the imaginary part of the action is preserved for the step size.
        The step size becomes updated accordingly.
        Further, only data that stems to be relevant for training, i.e. is not flowed to little and is not in neverland, is accepted.
    }\label{fig-train-data-alg}
\end{figure}
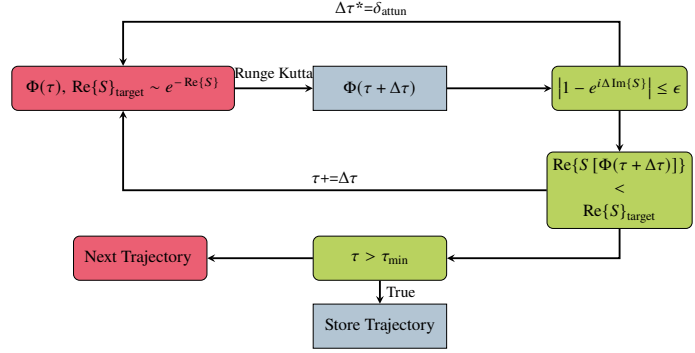

We train the parameters -- $w,w'\in\mathrm{Mat}_{\nicefrac{\abs{\Lambda}}{2}}\left[\Complexes\right], \ b,b' \in\Complexes^{\nicefrac{\abs{\Lambda}}{2}}$ per layer -- of the network using training data sets of flowed configurations~\cite{leveragingML}.
The training data is generated by integrating the holomorphic flow equation, starting from configurations sampled with a normal-distribution $\mathcal{N}\left(\mu=0,\sigma=\delta U\right)$ and mapped to the tangent plane, with a Runge Kutta and accepting only configurations that preserve the imaginary part of the action, are flowed enough, and are not in neverland
\footnote{
    By neverland we denote regions of infinite $\Re{S}$. Points where Lefshetz thimbles meet have diverging $\Re{S}$ and are not relevant for the training.
}.
A graphical representation of this can be found in figure~\ref{fig-train-data-alg}.
Once the data is gathered we initialize the network with random parameters $\sim U(0.01,0.01)$~\cite{pmlr-v9-glorot10a}, and train it using the Adam optimizer~\cite{kingma2014adam} with a learning rate of $10^{-3}$.
For the first iterations we use a plain $\mathrm{L}_2$-loss to measure the distance between the training data and the network output. 
After that a couple of iterations are trained by comparing the real part of the action plus the preservation of the imaginary part, $\mathrm{L}\left( \tilde{\Phi}, \NN{\Phi+\im\phi_0} \right) = \abs{\Delta \Re{S}} + \abs{1-e^{\im\Delta\Im{S}}}$.
Once a desired precision ($< 1e-4$) is reached, we perform a short HMC simulation, $\Ncfg = 1000$, and measure the statistical power $\abs{\Sigma}$.
If the statistical power is increased, we accept the new network parameters and continue the HMC simulation resulting in the observables presented in the next section.

\section{Results}
\subsection{Correlation Functions}

\begin{figure}[h]
    \centering
    \includegraphics[width=0.49\textwidth]{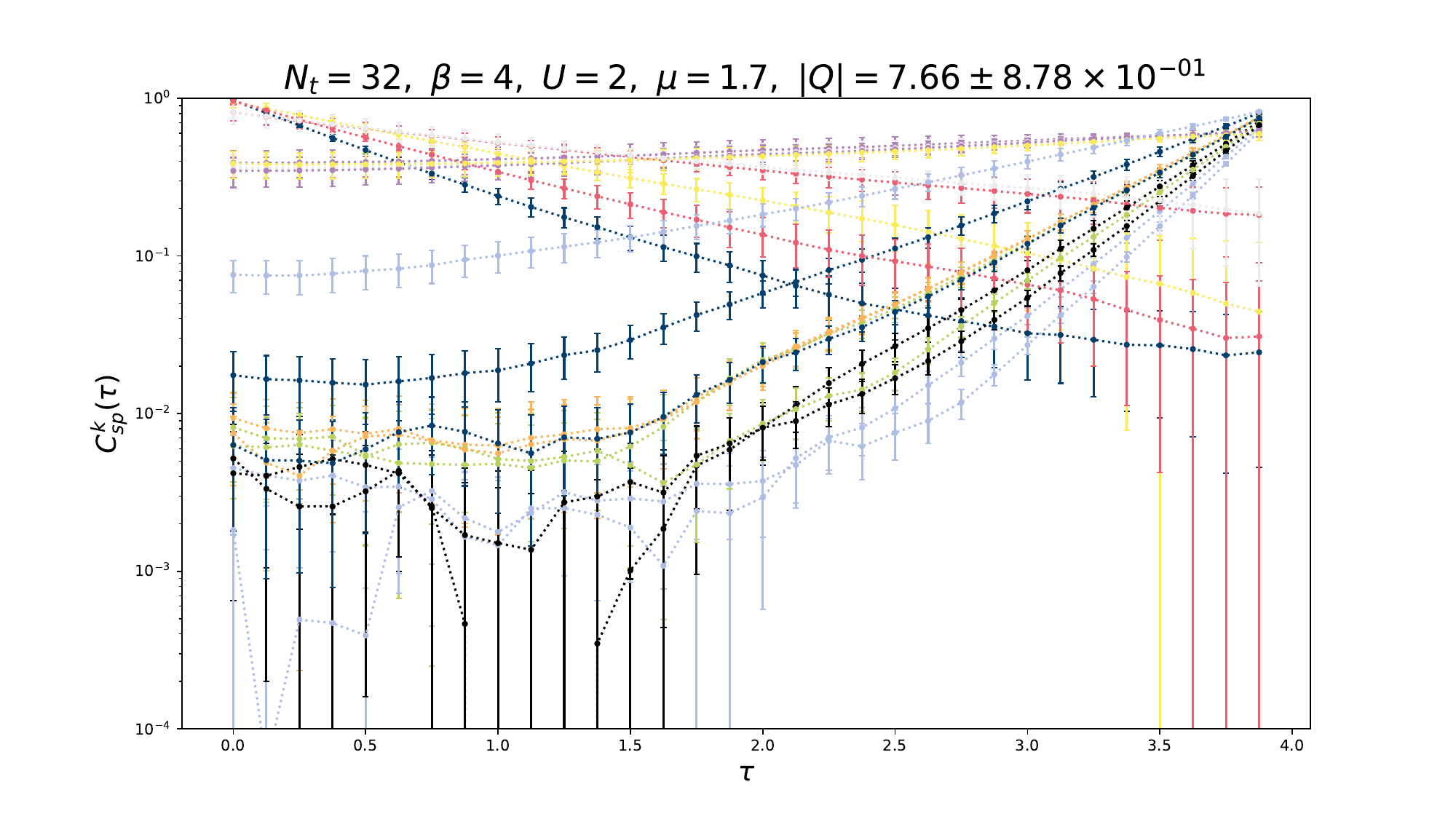} 
    \includegraphics[width=0.49\textwidth]{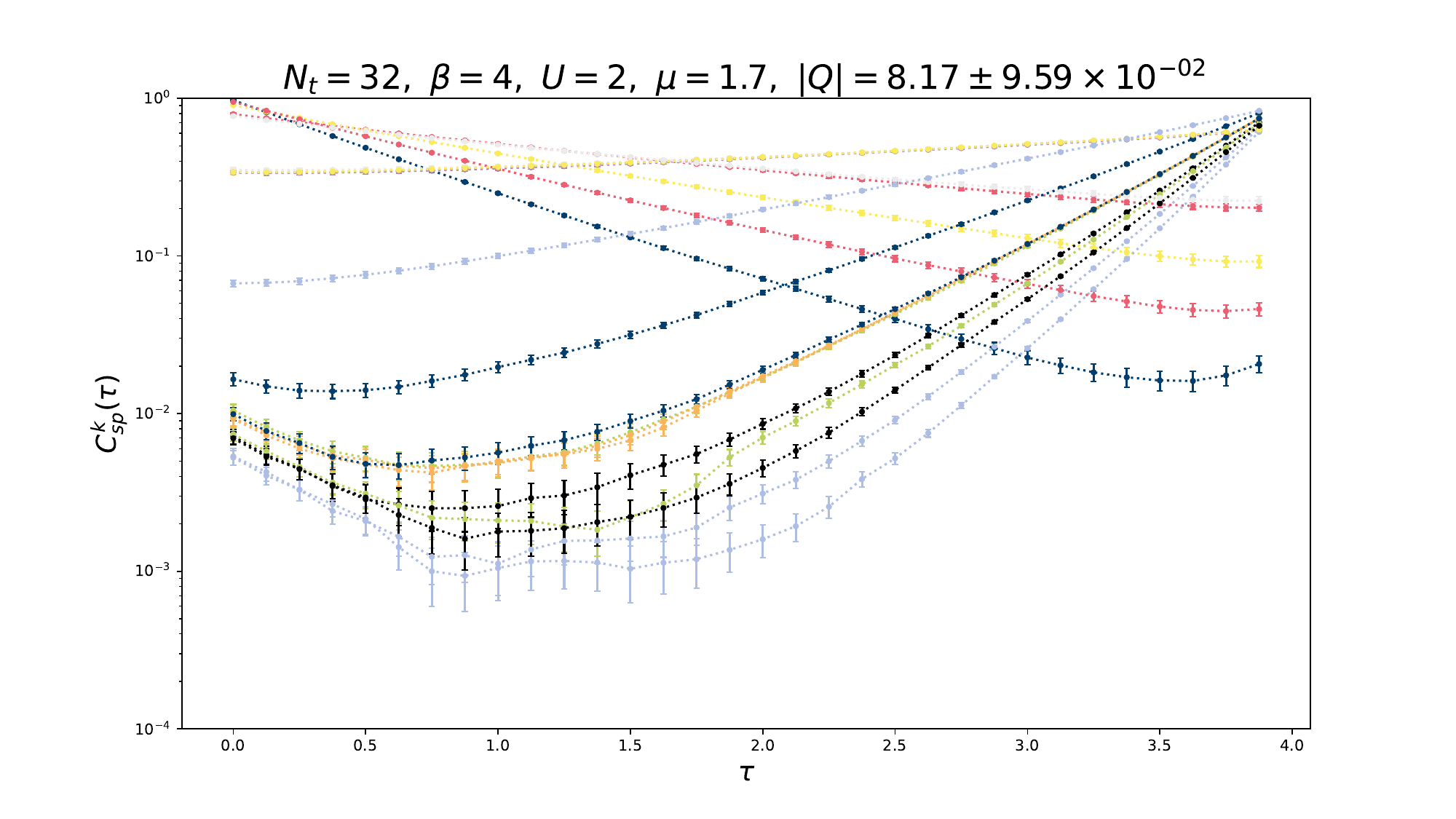}
    \caption{Perylene correlator for $U = 2,\,\mu=1.7$ simulated using HMC on the tangent plane (top) and machine learning guided HMC (bottom)}
    \label{fig-correlator}
\end{figure}

The basic building block of Monte-Carlo simulations for physical systems is the correlation function
\begin{equation}
    C^{x,y}(\tau) = \expval{ p_x(\tau) p^\dagger_y(0) } = \expval{ M^{-1}_{(\tau,x); (0,y)}\left[\Phi\, \vert \, K,\mu \right] }\label{eq-correlator},
\end{equation}
as it encodes the physics of a single excitation-annihilation process.
In the following we block diagonalize the correlation matrix using the symmetry eigenspaces of the ion lattice X
\begin{equation}
C^k(\tau) = \delta^{k,k'} \bar{\mathfrak{u}}_{k,x} C^{x,y}(\tau) \mathfrak{u}_{y,k'},
\end{equation}
where the unitary matrix $\mathfrak{u}$ is determined by the irreducible representations of the point group $D_{2h}$ of $X$. 
In principle, a further diagonalization within the irreducible blocks is required, however, we found that the diagonal terms are dominant in the cases we have studied thus far and so only take the diagonal correlators for study in these proceedings.
To classify the molecule and understand the entirety of its (single particle) spectrum a full parameter scan in the interaction strength and its doping needs to be performed as well as the continuum and zero temperature limit ($\beta,N_t \to \infty$).
In this preliminary results we will focus on the application of the neural network to the sign problem.
Hence, we fix $N_t = 32$ and $\beta = 4$, while the interaction and chemical potential strength are indicated at the specific results.
In Figures~\ref{fig-correlator} we compare correlator measurements from HMC on the tangent plane, on the left, and from machine learning enhanced HMC (MLHMC), on the right.
These are computed at fixed $U=2$ and $\mu = 1.7$.
As expected the MLHMC resolves the correlator much better than the HMC on the tangent plane.
This chemical potential is the first where the computational extra cost of the neural network can be justified.

\subsection{Charge}

\begin{figure}
    \centering
    \includegraphics[width=0.7\textwidth]{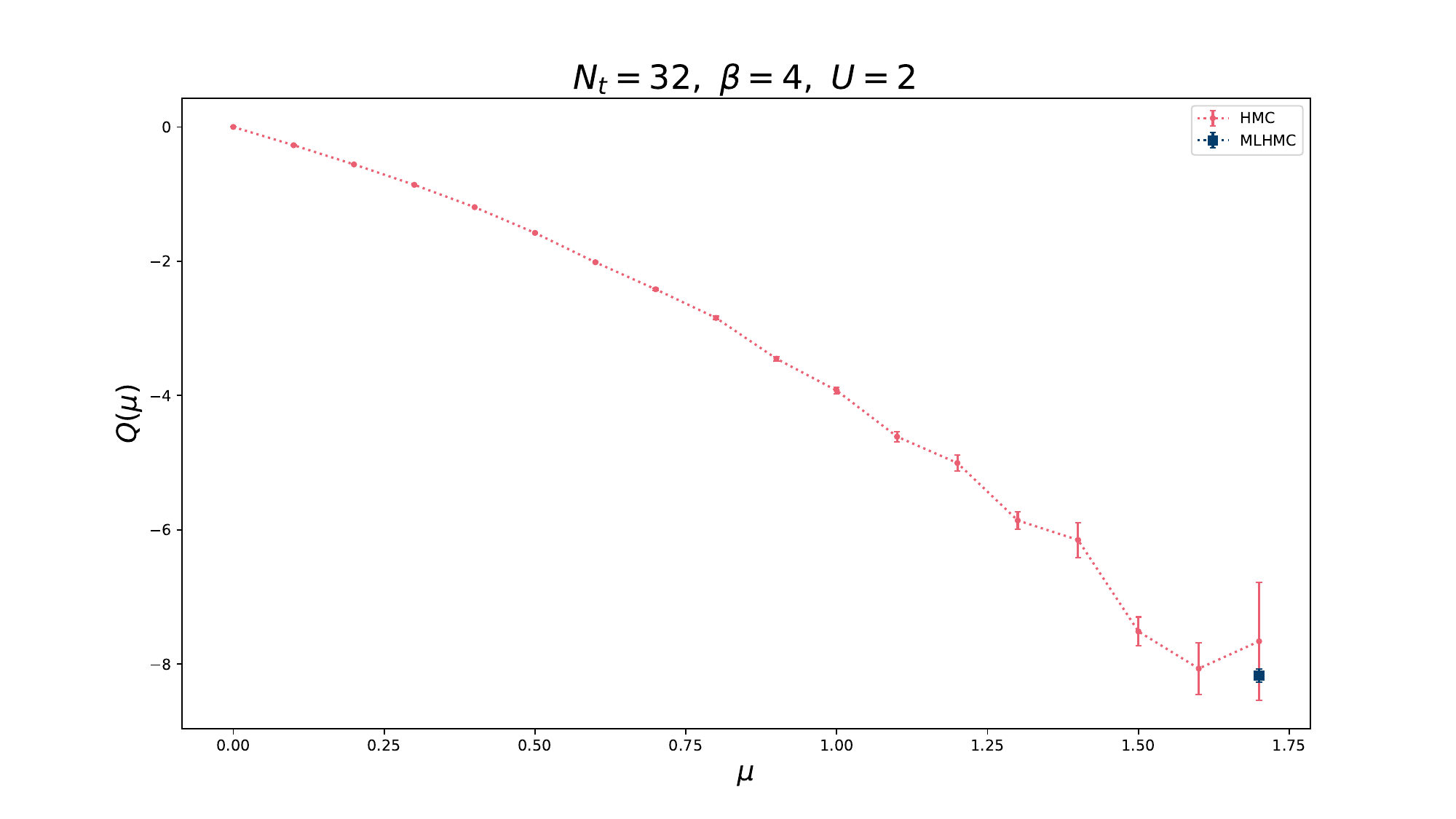}
    \caption{Perylene total charge $\mu$-scan at $U = 2$}
    \label{fig-charge-U2}
\end{figure}

The chemical potential is an effective description of the doping of the system. 
We translate this by measuring the total system charge
\begin{equation}
    Q = \sum_{x=0}^{N_x} \expval{ \rho_x } = \sum_{x=0}^{N_x} C^{x,x}(\tau=0).
\end{equation}
In Figure~\ref{fig-charge-U2} we present the total charge as a function of the chemical potential at fixed $U=2$.
The interaction strength is the same as for the correlators in Figure~\ref{fig-correlator}. 
We can identify, that the statistical error at $\mu = 1.7$ is much smaller for the MLHMC. 
At cold temperatures, $\beta \to \infty$, we expect a stepwise behaviour for integer charge changes at certain chemical potentials.
This is washed out here due to the small value of $\beta$.
To fully classify the doping of the molecule a zero-tempreature limit must be taken which will increase the use of the neural network as the sign problem exponentially becomes worse with increasing $\beta$.

\subsection{Effective Masses}

\begin{figure}
\centering
\begin{subfigure}[t]{0.5\textwidth}
    \centering
    \includegraphics[width=\textwidth]{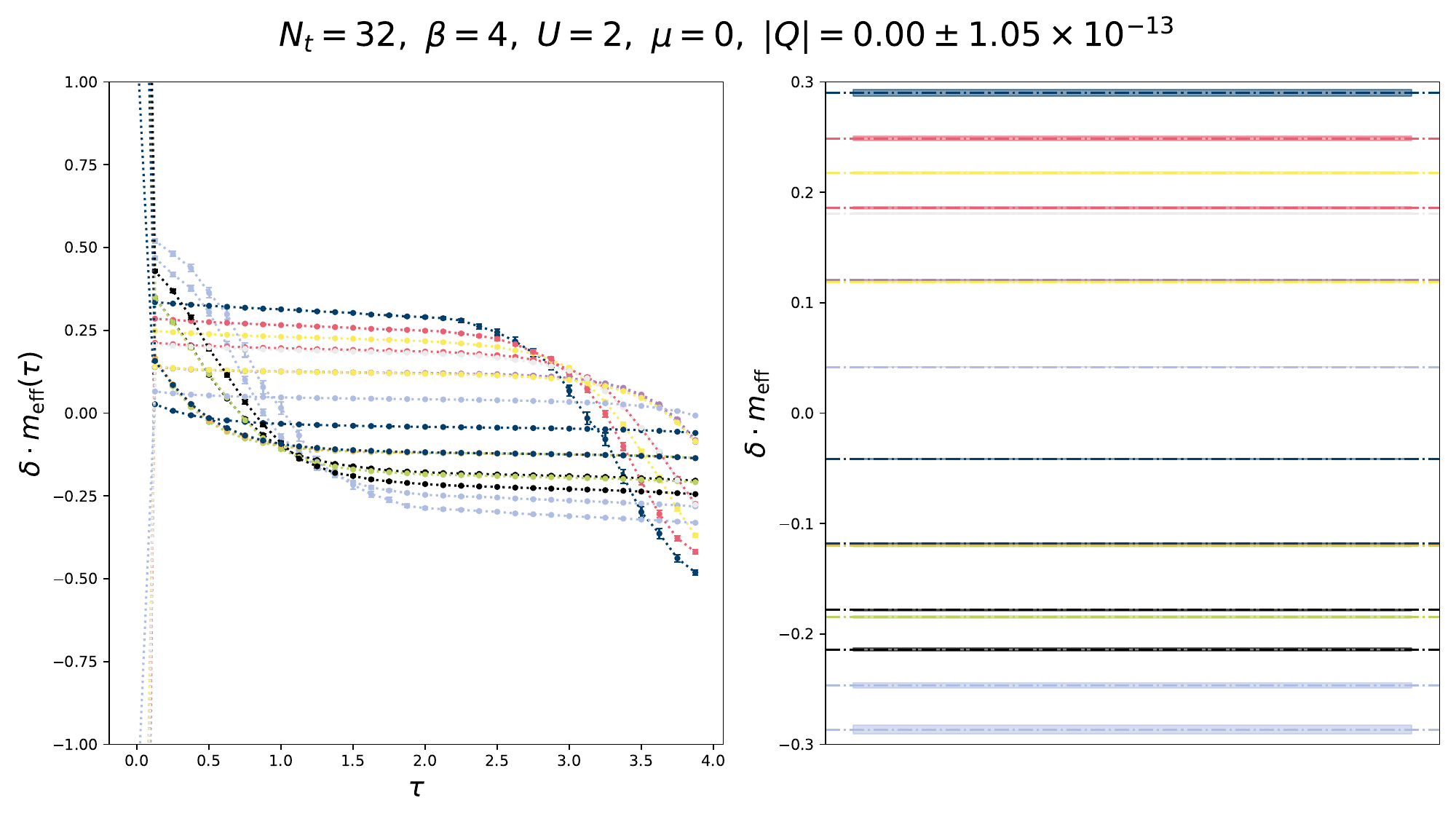}
    \caption{Perylene effective mass $U = 2,\,\mu=0$ (HMC)}
    \label{fig-meff-U2-mu0}
\end{subfigure}\hfill
\begin{subfigure}[t]{0.5\textwidth}
    \centering
    \includegraphics[width=\textwidth]{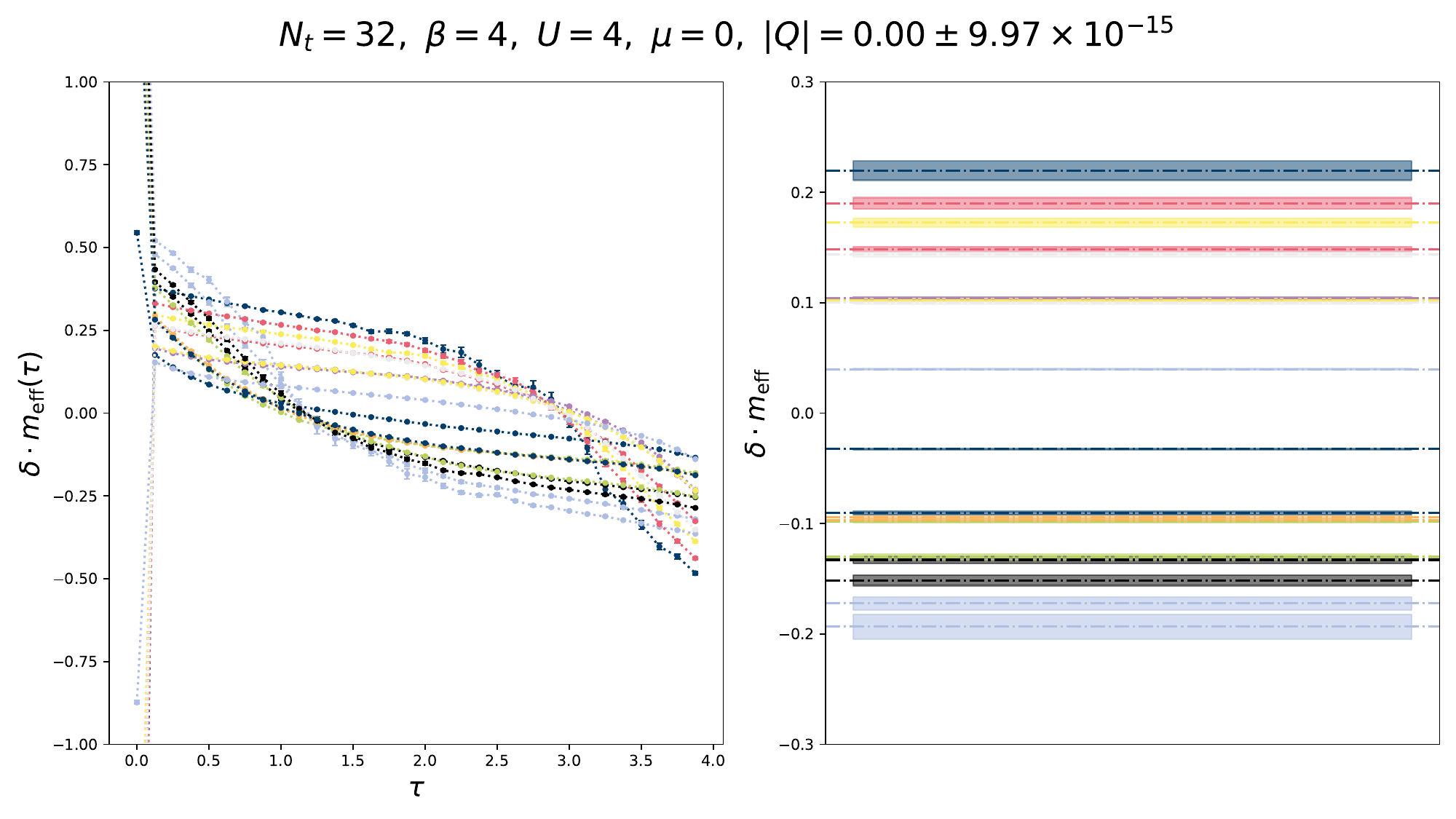}
    \caption{Perylene effective mass $U = 4,\,\mu=0$ (HMC)}
    \label{fig-meff-U4-mu0}
\end{subfigure}
\hfill
\begin{subfigure}[t]{0.5\textwidth}
    \centering
    \includegraphics[width=\textwidth]{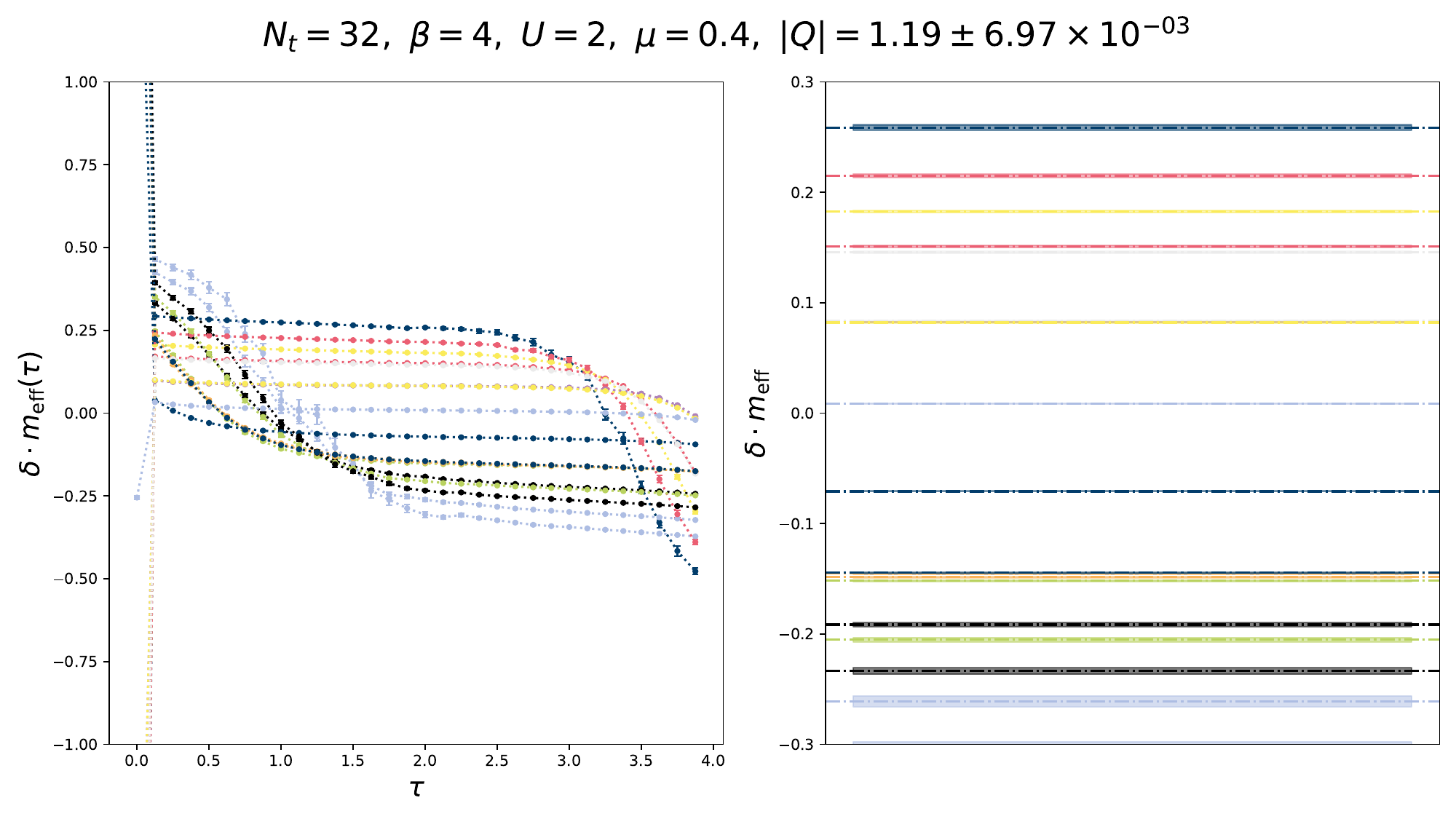}
    \caption{Perylene effective mass $U = 2,\,\mu=0.4$ (HMC)}
    \label{fig-meff-U2-mu0.4}
\end{subfigure}\hfill
\begin{subfigure}[t]{0.5\textwidth}
    \centering
    \includegraphics[width=\textwidth]{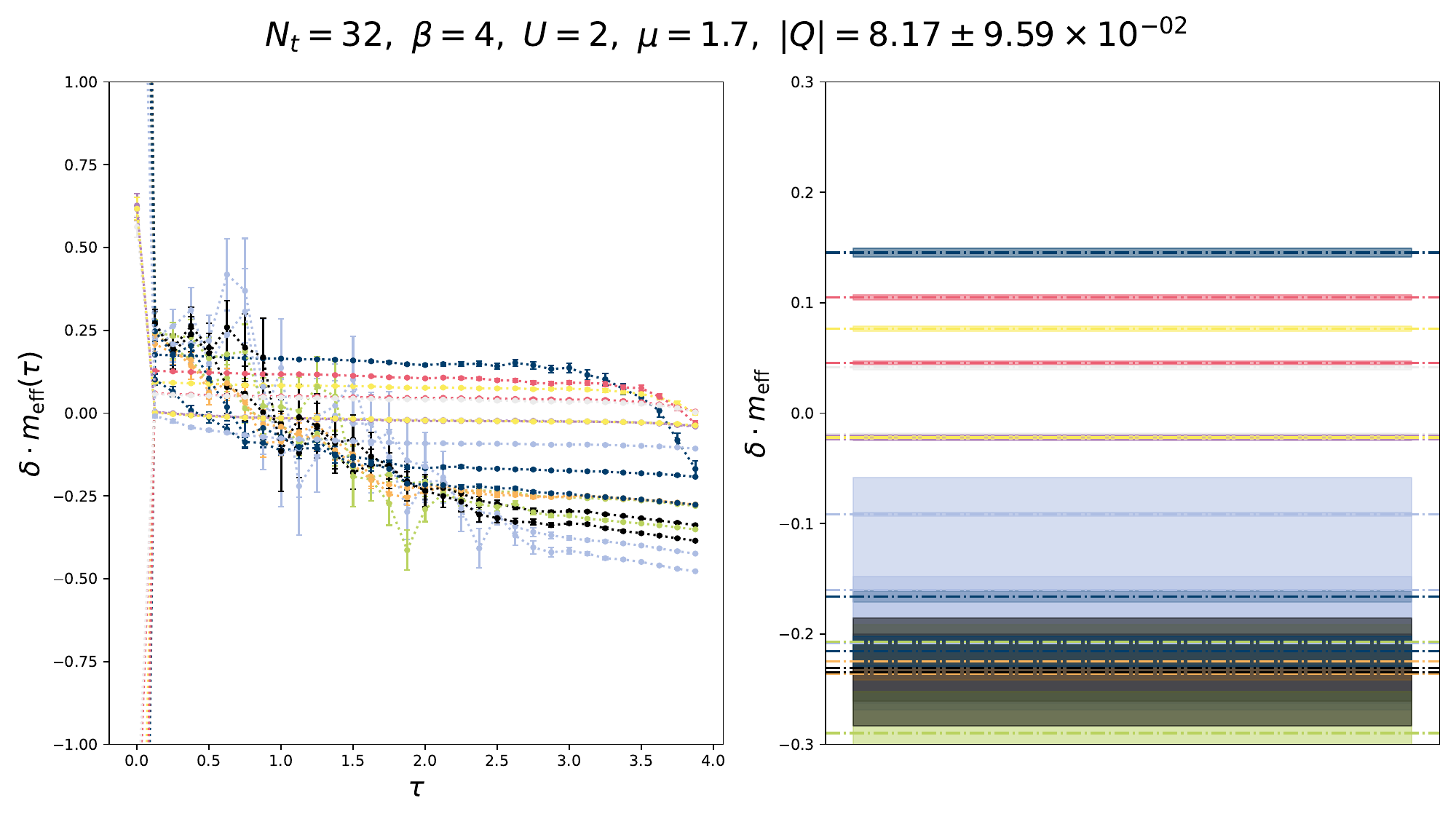}
    \caption{Perylene effective mass $U = 2,\,\mu=1.7$ (MLHMC)}
    \label{fig-meff-U2-mu1.7-MLHMC}
\end{subfigure}
\end{figure}

To study the molecule for various applications like energy production in organic solar cells it is required to get an understanding of the low lying energy spectrum of a particle-hole excitation.
At non-interacting systems $U=0$, this can be done analytically as the single-particle energy is given by the eigenvalues of the hopping matrix $E^k_{sp} = \epsilon^k$.
Doping simply shifts these  $E^k_{\nicefrac{sp}{sh}} = \epsilon^k \pm \mu$.
Further, if particles and holes do not interact the exciton energy is simply the sum $E^k = E^k_{sp} + E^k_{sh}$, which, noticeably, is independent of doping.
As soon as interactions are turned on this is not true anymore and the exciton energy needs to be measured explicitly involving computations of 4 point functions.
For this proceedings we take a first step into that direction and measure the single particle energy at various interaction strengths and chemical potentials.
In the low-temperature spectral decomposition 
$
C^{k}(\tau) = A_0 e^{-E_0^k \tau} + \order{e^{-E_1^k \tau}}
$
one expects ground state contributions in the large $\tau$ limit, defining the effective mass
\begin{equation}
    \left. m_\mathrm{eff}^k \right\vert_\tau = \frac{ \log C^k(\tau+\delta) - \log C^k(\tau)}{\delta}.
\end{equation}
The effective mass serves as a simple estimator for the energy $E^k_{sp} \approx\left. m_\mathrm{eff}^k \right\vert_{\tau = \nicefrac{\beta}{2}}$ that neglects excited states.
In figures~\ref{fig-meff-U2-mu0} to~\ref{fig-meff-U2-mu1.7-MLHMC} we present a selection of effective mass plots for two interaction strengths $U=2,4$ (\ref{fig-meff-U2-mu0},\ref{fig-meff-U4-mu0}) at zero chemical potential and two chemical potentials $\mu = 0.4,1.7$ at fixed $U=2$ (\ref{fig-meff-U2-mu0.4},\ref{fig-meff-U2-mu1.7-MLHMC}).
These plots suggest that the interaction strength has a great effect on the excited states, as the plateau becomes less pronounced, but only little on the ground state energy, i.e. the smallest positive energy value.
Further, we can identify that the gap between states opens as the interaction strength is increased.
Additionally, the change of chemical potential clearly shifts the ground state gap, which, together indicate a rich energy spectrum of the molecule.

\section{Summary}
The numerical sign problem affects Monte-Carlo simulations of various systems.
In this contribution we modelled the electronic structure of the $\text{sp}^2$-hybridized molecule $C_{20}H_{12}$ (perylene)  with a contact-interaction Hubbard model which is then simulated using 
the Hybird/Hamilton Monte-Carlo algorithm.
At finite chemical potential, a parameter to control doping, a sign problem appears which we tackle using reweighting and contour deformations.
Starting with simple deformations, the tangent plane, and continuing with neural networks at larger $\mu$ we are able to resolve the total system charge and single particle energy spectrum of perylene.
To limit thermal contamination of the energy states, we require that $\beta E_0 \gg 1$, where $E_0$ is the lowest single particle excitation energy in the system.  
In our case we have $\beta E_0 \sim 1.5$, which should be increased in future calculations.  
We plan to extend the parameter space and perform a more detailed analysis of the system, including the study of particle-hole excitations which are relevant for practical application of the molecule.

\begin{acknowledgments}
We gratefully acknowledge the computing time granted by the JARA Vergabegremium and provided on the JARA Partition part of the supercomputer JURECA at Forschungszentrum Jülich.
This work is supported by the MKW NRW under the funding code NW21-024-A.
Maria Dincă was funded by the German Academic Exchange Service (DAAD) in the program “RISE Germany”.
\end{acknowledgments}
 
\bibliographystyle{unsrt}

\end{document}